\documentclass[submission,copyright,creativecommons]{eptcs}
\providecommand{\event}{WoC 2015} 
\usepackage{breakurl}             

\title{From Push/Enter to Eval/Apply by Program Transformation}
\author{
Maciej Pir\'og \qquad\qquad Jeremy Gibbons
\institute{Department of Computer Science\\University of Oxford}
\email{maciej.pirog@cs.ox.ac.uk \qquad jeremy.gibbons@cs.ox.ak}
}
\def\titlerunning{From Push/Enter to Eval/Apply by Program Transformation}
\def\authorrunning{M. Pir\'og and J. Gibbons}

\usepackage{amsmath}
\usepackage{amsfonts}
\usepackage{xspace}
\usepackage{fancyvrb}

\newcommand{\lmulti}{$\lambda^{\!\mathrm{MULT}}$\xspace}
\newcommand{\pe}{PE\xspace}
\newcommand{\ea}{EA\xspace}
\newcommand{\Pe}{PE\xspace}
\newcommand{\Ea}{EA\xspace}
\newcommand{\papt}{\mathbf{pap}}
\newcommand{\conf}[2]{( #1 ,\, #2 )}
\newcommand{\abs}[2]{\lambda \langle #1 \rangle . #2}
\newcommand{\app}[2]{#1 \, \langle #2 \rangle }
\newcommand{\gsep}{\ \, | \ \,}
\newcommand{\mtrans}{\Rightarrow}
\newcommand{\cons}{{:}}
\newcommand{\emptystack}{\varepsilon}
\newcommand{\subs}[1]{[ #1 ]}
\newcommand{\sub}[2]{#2/#1}
\newcommand{\cont}[1]{\langle #1 \rangle}
\newcommand{\kpap}[4]{\papt( #1 , #3 , #4 )}

\newcommand{\rulename}[1]{$\mathsf{#1}$}

\begin{document}
\maketitle

\begin{abstract}
Push/enter and eval/apply are two calling conventions used in implementations of functional languages. In this paper, we explore the following observation: when considering functions with multiple arguments, the stack under the push/enter and eval/apply conventions behaves similarly to two particular implementations of the list datatype: the regular cons-list and a form of lists with lazy concatenation respectively. Along the lines of Danvy~\textit{et al.}'s functional correspondence between definitional interpreters and abstract machines, we use this observation to transform an abstract machine that implements push/enter into an abstract machine that implements eval/apply. We show that our method is flexible enough to transform the push/enter Spineless Tagless G-machine (which is the semantic core of the GHC Haskell compiler) into its eval/apply variant.
\end{abstract}

\section{Introduction}

There are two standard calling conventions used to efficiently compile curried multi-argument functions in higher-order languages: \emph{push/enter} (\pe) and \emph{eval/apply} (\ea). With the \pe convention, the caller pushes the arguments on the stack, and jumps to the function body. It is the responsibility of the function to find its arguments, when they are needed, on the stack. With the \ea convention, the caller first evaluates the function to a normal form, from which it can read the number and kinds of arguments the function expects, and then it calls the function body with the right arguments. The difference between the two calling conventions is thoroughly discussed by Marlow and Peyton Jones~\cite{DBLP:journals/jfp/MarlowJ06}.

The terms \pe and \ea are also used in the literature to describe abstract machines, which deal directly with program expressions, and thus do not distinguish between `code' of a function and a caller. Roughly speaking, a \pe machine keeps on its stack some kind of \emph{resources}, which can be accessed whenever needed. In case of curried functions, they are arguments, which can be freely used by a function body. An \ea machine keeps on its stack \emph{continuations}, which are very often synonymous with evaluation contexts~\cite{citeulike:4128, BRICS-RS-04-26}. Such a machine has two kinds of configurations: `eval' configurations evaluate (sub)expressions independently of the context, while `apply' configurations construct a new expression from the obtained value and the next continuation.

Encouraged by the fact that some machines (such as the Spineless Tagless G-machine (STG)~\cite{DBLP:journals/jfp/MarlowJ06} used in the Glasgow Haskell Compiler) come in two versions realising the two conventions, one may suspect that the calling convention of a machine is orthogonal to the other aspects of its computational model. If so, is there a generic method of relating two such incarnations of a single machine? In this article, we give a partial answer to this question: we present a semi-mechanical method of derivation of an \ea machine from its \pe counterpart.

Our method is based on the correspondence between interpreters written in a functional language and abstract machines, studied extensively by Danvy \textit{et al.}~\cite{DBLP:conf/ppdp/AgerBDM03}. The key observation
is that tail-recursive functions are similar in structure to state-transition systems, and thus tail-recursive interpreters of programming calculi correspond to abstract machines. Moreover, a flat tail-recursive structure can always be obtained via CPS transformation followed by defunctionalisation of the resulting continuations. This method can be used to transform between different abstract machines, as long as we can encode the initial abstract machine as a functional program:
\begin{center}
\begin{tabular}{c}abstract\\machine\end{tabular}
$\xrightarrow{\text{encoding}}$
\begin{tabular}{c}functional\\program\end{tabular}
$\xrightarrow{\text{program transformation}}$
\begin{tabular}{c}tail-recursive\\functional\\program\end{tabular}
$\xrightarrow{\text{decoding}}$
\begin{tabular}{c}abstract\\
machine
\end{tabular}
\end{center}
Our method proposes a new tool for the `encoding' part, as there are multiple choices of how one represents an abstract machine as a functional program. In particular, one can use different data structures to represent the type of the stack of the original machine. In detail, our derivation consists of the following steps:

\begin{itemize}
\item We start with a \Pe machine, which is a slight generalisation of a machine proposed by Krivine~\cite{DBLP:journals/lisp/Krivine07}. It normalises terms to weak-head normal form using the call-by-name evaluation strategy. Then, we give a (big-step~\cite{DBLP:journals/ipl/DanvyM08}) encoding of this machine in Haskell. (We do not use any Haskell-specific features, and the choice of Haskell over any other functional language is in this case arbitrary. The `lazy list concatenation' that we use in this paper is unrelated to the lazy semantics of Haskell.)

\item We then perform semantics-preserving program transformations on this encoding. They are purely syntactic, that is they are not directed by any semantic understanding of the machine. In detail, we implement the stack using a list with lazy concatenation, and then apply the CPS transformation to reify the recursive calls of the operations that work on the stack as transitions of the machine. 

\item Finally, we decode the resulting \ea machine from the program obtained by these transformations.
\end{itemize}

There are some strong similarities between the resulting machines and the application-related rules of the two versions of the STG machine~\cite{DBLP:journals/jfp/MarlowJ06} used in the Haskell GHC compiler. Thus our method proves to be useful even when dealing with real-life implementations.

\section{The Krivine machine for a language with multi-argument binders}

We first define \lmulti, a version of the lambda calculus in which a $\lambda$-abstraction can bind more than one variable at a time, and in which an expression can be applied to a tuple of arguments. Its syntax is given by the following grammar, where $e, e_0, \ldots$ stand for expressions, $x, x_1, \ldots$ denote variables, and $\langle a_1 \ldots a_n \rangle$ is a non-empty tuple containing elements $a_1, \ldots, a_n$:
\begin{equation}\notag
e\ {:}{:}{=}\ x
  \gsep \app{e_0}{e_1 \ldots e_n}
  \gsep \abs{x_1 \ldots x_n}{e}
\end{equation}
Intuitively, an expression $\app{e_0}{e_1 \ldots e_n}$ roughly corresponds to $e_0 e_1 \cdots e_n$ in the standard lambda calculus, while $\abs{x_1 \ldots x_n}{e}$ corresponds to $\lambda x_1. \ldots \lambda x_n. e$.
Note that the \lmulti calculus is different from the one obtained by simply adding finite products to the standard lambda calculus and treating $\lambda$-abstractions as uncurried functions: for example, the term $\app{\app{(\abs{x_1 \ldots x_4}{e})}{x_1 \, x_2}}{x_3 \, x_4}$ has an `arity' mismatch in the latter calculus.

We give an operational semantics to \lmulti in terms of a \pe abstract machine, which normalises a given expression to weak-head normal form using the call-by-name evaluation strategy. The machine is a natural extension of an abstract machine proposed by Krivine~\cite{DBLP:journals/lisp/Krivine07} (see also Biernacka and Danvy~\cite{DBLP:journals/tocl/BiernackaD07}), which 
was originally defined for a language with multi-argument binders, but with applications limited to a single argument (and not a tuple of arguments). Each machine configuration is a pair $\conf{e}{s}$ consisting of an expression and a stack of expressions. We write `${\cons}$' for the stack constructor (the `push' operation), and $\emptystack$ for the empty stack. We write $e \subs{\sub{x_1}{e_1}, \ldots , \sub{x_n}{e_n}}$ to denote an expression $e$ in which expressions $e_1, \ldots, e_n$  are substituted for variables $x_1, \ldots , x_n$ respectively. The transition rules are as follows:
\begin{align}
\tag{\rulename{K\text{-}APP}}
\conf{\app{e_0}{e_1 \ldots e_n}}{s} &\mtrans \conf{e_0}{e_1 \cons \cdots \cons e_n \cons s} \label{m:k2}
\\
\tag{\rulename{K\text{-}FUN}}
\conf{\abs{x_1 \ldots x_n}{e}}{e_1 \cons \cdots \cons e_n \cons s} &\mtrans \conf{e \subs{\sub{x_1}{e_1} , \ldots , \sub{x_n}{e_n}}}{s} \label{m:k3}
\end{align}
The initial configuration of the machine for an expression $e$ is $\conf{e}{\emptystack}$. The transition \ref{m:k2} evaluates applications: it pushes the arguments on the stack and continues with the head of the application. The transition \ref{m:k3} deals with abstractions: if the abstraction needs $n$ arguments and there are at least $n$ expressions on the stack, the machine continues with the body of the abstraction, with the appropriate variables substituted. The machine halts when there are no transitions that match the left-hand sides of \ref{m:k2} or~\ref{m:k3}, that is, when trying to evaluate an application for which there are not enough arguments on the stack or when trying to evaluate a free variable.

\section{Haskell implementation}

Now, we present a deep embedding of \lmulti and the machine in Haskell. The definition of terms is as follows:
\begin{Verbatim}[frame=single, framerule=0mm, framesep=0.5em, xleftmargin=2em]
type Identifier = ...
data Term = Var Identifier
          | App Term [Term]
          | Fun [Identifier] Term

subst :: [(Term, Identifier)] -> Term -> Term
subst = ...
\end{Verbatim}
Neither the choice of the type of identifiers nor any concrete implementation of \verb+subst+ affect the derivation in any way.
The Haskell encoding of the machine uses an abstract datatype representing the stack of the machine. It has two operations:
\begin{Verbatim}[frame=single, framerule=0mm, framesep=0.5em, xleftmargin=2em]
type Stack a = ...
push :: [a] -> Stack a -> Stack a
pop  :: Int -> Stack a -> Maybe ([a], Stack a)
\end{Verbatim}
In the above, the function \verb+push+ places a tuple of values on top of the stack. The function \verb+pop n s+ attempts to extract the first \verb+n+ values from the top of the stack \verb+s+. If the stack contains too few elements, \verb+pop+ returns \verb+Nothing+.
The most obvious implementation of this interface uses the list datatype:
\begin{Verbatim}[frame=single, framerule=0mm, framesep=0.5em, xleftmargin=2em]
type Stack a = [a]

push :: [a] -> Stack a -> Stack a
push = (++)

pop  :: Int -> Stack a -> Maybe ([a], Stack a)
pop n xs | length xs >= n = Just (take n xs, drop n xs)
         | otherwise      = Nothing
\end{Verbatim}
The encoding of the transition rules is straightforward. We include the cases in which the machine halts:
\begin{Verbatim}[frame=single, framerule=0mm, framesep=0.5em, xleftmargin=2em]
eval :: Term -> Stack Term -> (Term, Stack Term)
eval (App t ts) s = eval t (push ts s)                          -- K-APP
eval (Fun xs t) s =
  case pop (length xs) s of
    Just (ts1, ts2) -> eval (subst (zip ts1 xs) t) ts2          -- K-FUN
    Nothing         -> (Fun xs t, s)                            -- halt
eval (Var i)    s = (Var i, s)                                  -- halt
\end{Verbatim}

\section{Deriving a \Pe machine}

Now, we transform the program described in the previous section. We proceed in three stages. First, we choose a different implementation of the stack datatype and the two associated operations. Then we perform CPS transformation, which reifies the recursive calls in \verb+pop+ as steps of execution of the abstract machine. Finally, to eliminate the higher-order functions that arise from CPS transformation, we perform inlining.

\subsection{Lazy concatenation}

Another possible implementation of the \verb+Stack+ datatype is a list with lazy concatenation. It is defined as a list of lists, so that the \verb+push+ operation can be defined simply as consing the first argument to the front of the structure:
\begin{Verbatim}[frame=single, framerule=0mm, framesep=0.5em, xleftmargin=2em]
type Stack a = [[a]]

push :: [a] -> Stack a -> Stack a
push xs xss = xs : xss

pop :: Int -> Stack a -> Maybe ([a], Stack a)
pop n ys = pop' [] n ys

pop' :: [a] -> Int -> Stack a -> Maybe ([a], Stack a)
pop' acc n ys
  | m == n                 = Just (acc, ys)
  | m > n                  = Just (take n acc, drop n acc : ys)
  | m < n && length ys > 0 = pop' (acc ++ head ys) n (tail ys)
  | otherwise              = Nothing
  where m = length acc
\end{Verbatim}
The definition of \verb+pop+ uses an auxiliary function \verb+pop'+, which is tail-recursive. We maintain the invariant that there are no empty inner lists. That is why we include the separate case \verb+m == n+ to make sure that \verb+pop'+ does not leave an empty list at the front of the structure.

We use these definitions in our encoding of the machine. Note that we do not do that for performance reasons, because we do not intend to ever execute these programs. We use Haskell as a metalanguage to derive one abstract machine from another abstract machine. The equivalence of the programs guarantees the equivalence of the machines. We still use regular cons-lists to represent tuples, hence standard list functions \verb+length+ and \verb+zip+.

\subsection{CPS translation}

Since \verb+pop'+ is tail-recursive, we can see its definition as an encoding of yet another abstract machine. Our intention is to fuse the two machines together. Thus, \verb+eval+ cannot treat \verb+pop+ as an atomic operation any longer. We need an explicit transfer of control: \verb+eval+ calls \verb+pop+, which calls \verb+eval+ back instead of just returning a value. We can easily achieve this behaviour with the call-by-value CPS translation of \verb+pop+ and \verb+pop'+.
We obtain functions \verb+popCPS+ and \verb+pop'CPS+, with the property \verb+popCPS n ys k = k (pop n ys)+ and \verb+pop'CPS acc n ys k = k (pop' acc n ys)+ for any continuation \verb+k+, that is, a function of the type \verb+k :: (Maybe ([a], Stack a) -> o)+. From this, we easily obtain that \verb+evalCPS+ given below is extensionally equal to \verb+eval+.

\begin{Verbatim}[frame=single, framerule=0mm, framesep=0.5em, xleftmargin=2em]
popCPS :: Int -> Stack a -> (Maybe ([a], Stack a) -> o) -> o
popCPS n ys k = pop'CPS [] n ys k

pop'CPS :: [a] -> Int -> Stack a -> (Maybe ([a], Stack a) -> o) -> o
pop'CPS acc n ys k
  | m == n                 = k (Just (acc, ys))
  | m > n                  = k (Just (take n acc, drop n acc : ys))
  | m < n && length ys > 0 = pop'CPS (acc ++ head ys) n (tail ys) k
  | otherwise              = k Nothing
  where m = length acc

evalCPS :: Term -> Stack Term -> (Term, Stack Term)
evalCPS (App t ts) s = evalCPS t (push ts s)
evalCPS (Fun xs t) s = popCPS (length xs) s aux
 where
  aux (Just (ts1, ts2)) = evalCPS (subst (zip ts1 xs) t) ts2
  aux Nothing           = (Fun xs t, s)
evalCPS (Var i)    s = (Var i, s)
\end{Verbatim}

\subsection{Inlining}

We observe that in the entire program the function \verb+popCPS+ is called in one place only, and its continuation \verb+aux+ is not modified throughout the recursive calls of \verb+pop'CPS+. So, we know what the continuation inside \verb+popCPS+ and \verb+pop'CPS+ is, up to its free variables. We can inline the body of the continuation, and supply values for its free variables as arguments of the \verb+popCPS+ operation. In detail, we first lambda-lift \verb+aux+. As a result, it becomes a separate function:
\begin{Verbatim}[frame=single, framerule=0mm, framesep=0.5em, xleftmargin=2em]
evalCPS2 :: Term -> Stack Term -> (Term, Stack Term)
evalCPS2 (App t ts) s = evalCPS2 t (push ts s)
evalCPS2 (Fun xs t) s = popCPS (length xs) s (aux2 xs t s)
evalCPS2 (Var i)    s = (Var i, s)

aux2 :: [Identifier] -> Term -> Stack Term -> Maybe ([Term], Stack Term)
                                                     -> (Term, Stack Term)
aux2 xs t s (Just (ts1, ts2)) = evalCPS2 (subst (zip ts1 xs) t) ts2
aux2 xs t s Nothing           = (Fun xs t, s)
\end{Verbatim}
Second, we inline the definitions of \verb+popCPS+ and \verb+push+ in \verb+evalCPS2+. Additionally, we define a function \verb+pop'In+ with the property that \verb+pop'CPS vs n s (aux xs t s) = pop'In vs n t xs s+. Applying this equality in the definition of \verb+evalCPS2+, we obtain the following:
\begin{Verbatim}[frame=single, framerule=0mm, framesep=0.5em, xleftmargin=2em]
pop'In :: [Term] -> Int -> Term -> [Identifier] -> Stack Term
                                                    -> (Term, Stack Term)
pop'In acc n t xs ys
  | m == n = evalIn (subst (zip acc xs) t) ys
  | m > n  = evalIn (subst (zip (take n acc) xs) t) (drop n acc : ys)
  | m < n && length ys > 0 = pop'In (acc ++ head ys) n t xs (tail ys)
  | otherwise              = (Fun xs t, [])
  where m = length acc

evalIn :: Term -> Stack Term -> (Term, Stack Term)
evalIn (App t ts) s = evalIn t (ts : s)
evalIn (Fun xs t) s = pop'In [] (length xs) t xs s
evalIn (Var i)    s = (Var i, s)
\end{Verbatim}

Now, we are almost ready to decode the final machine. The very last observation is that the argument \verb+n+ is redundant in \verb+pop'In+: it is always equal to the length of \verb+xs+, since they are equal in the call in \verb+evalIn+, and they are not changed throughout the recursive calls. Therefore, we can eliminate it in the decoding, and compare the length of the accumulator to the length of \verb+xs+. So, the Haskell encoding becomes as follows, in which we have \verb+pop'In2 acc t xs ys = pop'In acc (length xs) t xs ys+.
\begin{Verbatim}[frame=single, framerule=0mm, framesep=0.5em, xleftmargin=2em]
pop'In2 :: [Term] -> Term -> [Identifier] -> Stack Term
                                                  -> (Term, Stack Term)
pop'In2 acc t xs ys
  | m == n = evalIn2 (subst (zip acc xs) t) ys
  | m > n  = evalIn2 (subst (zip (take n acc) xs) t) (drop n acc : ys)
  | m < n && length ys > 0 = pop'In2 (acc ++ head ys) t xs (tail ys)
  | otherwise              = (Fun xs t, [])
  where m = length acc
        n = length xs

evalIn2 :: Term -> Stack Term -> (Term, Stack Term)
evalIn2 (App t ts) s = evalIn2 t (ts : s)
evalIn2 (Fun xs t) s = pop'In2 [] t xs s
evalIn2 (Var i)    s = (Var i, s)
\end{Verbatim}

\begin{figure*}[t]
\newcommand{\Mtrans}{\Rightarrow \ \ }
\begin{align}
\tag{\rulename{E\text{-}APP}}\label{m:e2}
&\conf{\app{e_0}{e_1 \ldots e_k}}{s} \\ \notag
\Mtrans &\conf{e_0}{\cont{e_1 \ldots e_k} \cons s} 
\\[7 pt]
\tag{\rulename{E\text{-}FUN}}\label{m:e3}
&\conf{\abs{x_1 \ldots x_n}{e}}{s} \\ \notag
\Mtrans &\conf{\kpap{\emptystack}{n}{e}{x_1 \ldots x_n}}{s}
\\[7 pt]
\tag{\rulename{A\text{-}EQ}}\label{m:e4}
&\conf{\kpap{e_1 \cons \cdots \cons e_n \cons \emptystack}{n}{e}{x_1 \ldots x_n}}{s} \\ \notag
\Mtrans &\conf{e \subs{\sub{x_1}{e_1} , \ldots , \sub{x_n}{e_n}}}{s}
\\[7 pt]
\tag{\rulename{A\text{-}GT}}\label{m:e5}
&\conf{\kpap{e_1 \cons \cdots \cons e_k \cons \emptystack}{n}{e}{x_1 \ldots x_n}}{s} \\ \notag
\Mtrans &\conf{e \subs{\sub{x_1}{e_1} , \ldots , \sub{x_n}{e_n}}}{\cont{e_{n+1} \ldots e_k} \cons s}
\text{, where $k > n$}
\\[7 pt]
\tag{\rulename{A\text{-}LT}}\label{m:e6}
&\conf{\kpap{e_1 \cons \cdots \cons e_k \cons \emptystack}{n}{e}{x_1 \ldots x_n}}{\cont{f_1 \ldots f_m} \cons s} \\ \notag
\Mtrans \notag  &\conf{\kpap{e_1 \cons \cdots \cons e_k \cons f_1 \cons \cdots \cons f_m \cons \emptystack}{n}{e}{x_1 \ldots x_n}}{s}
\text{, where $k < n$}
\end{align}
\caption{The eval/apply abstract machine}
\label{fig:e}
\end{figure*}

We are left with two tail- and mutually-recursive functions. Seen as transition rules, they represent two types of configuration of an abstract machine: \verb+evalIn2+ represents `eval' configurations with an expression and a stack, while \verb+pop'In2+ represents `apply' configurations, which we denote $\papt$ (short for `partial application'). The latter consists of a list of accumulated actual arguments (\verb+acc+), the body of the abstraction~(\verb+t+), a tuple of formal arguments (\verb+xs+), and the stack (\verb+ys+). The transitions of the machine are shown in Figure~\ref{fig:e}.

Intuitively, a $\papt$ configuration stores an abstraction and a small stack of accumulated actual arguments. When there are enough arguments, the application is performed (\ref{m:e4} and \ref{m:e5}). When there are too few arguments, the small stack is extended with the arguments from the top frame of the main stack (\ref{m:e6}).

\section{The applicative fragment of the STG machine}

\begin{figure*}[t]
\centering
\newcommand{\Mtrans}{\Rightarrow \ \ }
\begin{align}
\tag{\rulename{STG\text{-}TCALL}}\label{m:s2}
&\conf{\app{e_0}{e_1 \ldots e_k}}{s}\\
\Mtrans &\conf{e_0}{\cont{e_1 \ldots e_k} \cons s} \notag
\text{, where $e_0$ is not an abstraction}
\\[7 pt]
\tag{\rulename{STG\text{-}EXACT}}\label{m:s3}
&\conf{\app{(\abs{x_1 \ldots x_n}{e})}{e_1 \ldots e_n}}{s} \\
\Mtrans &\conf{e \subs{\sub{x_1}{e_1} , \ldots , \sub{x_n}{e_n}}}{s} \notag
\\[7 pt]
\tag{\rulename{STG\text{-}CALLK}}\label{m:s4}
&\conf{\app{(\abs{x_1 \ldots x_n}{e})}{e_1 \ldots e_k}}{s} \\
\Mtrans &\notag \hfill \conf{e \subs{\sub{x_1}{e_1} , \ldots , \sub{x_n}{e_n}}}{\cont{e_{n+1}\ldots e_k} \cons s} \notag
\text{, where $k > n$}
\\[7pt]
\tag{\rulename{STG\text{-}PAP2}}\label{m:s5}
&\conf{\app{(\abs{x_1 \ldots x_n}{e})}{e_1 \ldots e_k}}{s} \\
\Mtrans &\notag \conf{\kpap{e_1 \cons \cdots \cons e_k \cons \emptystack}{n}{e}{x_1 \ldots x_n}}{s} \notag
\text{, where $k < n$}
\\[7pt]
\tag{\rulename{STG\text{-}RETFUN}}\label{m:s6}
&\conf{\abs{x_1\ldots x_n}{e}}{\cont{e_1 \ldots e_k} \cons s} \\
\Mtrans &\conf{\app{(\abs{x_1\ldots x_n}{e})}{e_1 \ldots e_k}}{s} \notag
%
\\[7pt]
\tag{\rulename{STG\text{-}PCALL}}\label{m:s7}
&\conf{\kpap{e_1 \cons \cdots \cons e_n \cons \emptystack}{n}{e}{x_1 \ldots x_k}}{\cont{f_1 \ldots f_m} \cons s} \\
\Mtrans &\notag  \conf{\app{(\abs{x_1 \ldots x_k}{e})}{e_1 \ldots e_n \, f_1 \ldots f_m}}{s}
\end{align}
\caption{The STG-like abstract machine}
\label{fig:s}
\end{figure*}

The machine that we arrived at is the applicative fragment of the \ea STG machine (as introduced by Marlow and Peyton Jones~\cite{DBLP:journals/jfp/MarlowJ06}) in disguise. We only need to slightly rearrange the rules, as shown in Figure~\ref{fig:s}. First, if $e_0$ in \ref{m:e2} is an abstraction, the transition is always followed by \ref{m:e3}, which can be followed only by \ref{m:e6} (since $n$ in \ref{m:e3} is greater than 0), and then one of the rules \ref{m:e4}, \ref{m:e5}, or \ref{m:e6}. We fuse these three possible execution paths into three rules: \ref{m:s2}, \ref{m:s3}, and \ref{m:s4}. We can reuse them to deal with abstractions when there are some arguments on the stack, by constructing an application (\ref{m:s5}), so that a path \ref{m:e3}$\Rightarrow$\ref{m:e6}$\Rightarrow$(\ref{m:e4}, \ref{m:e5}, or \ref{m:e6}) becomes \ref{m:s5}$\Rightarrow$(\ref{m:s2}, \ref{m:s3}, or \ref{m:s4}). We proceed similarly with the $\papt$ configurations (\ref{m:s6}), so that transitions \ref{m:e4}, \ref{m:e5}, and \ref{m:e6} become \ref{m:s6}$\Rightarrow$\ref{m:s2}, \ref{m:s6}$\Rightarrow$\ref{m:s3}, and \ref{m:s6}$\Rightarrow$\ref{m:s4} respectively. Note that the rearranging of the rules can be done also on the level of Haskell programs by means of inlining and expanding of the definitions.

Marlow and Peyton Jones~\cite{DBLP:journals/jfp/MarlowJ06} introduced two versions of the STG machine: push/enter and eval/apply, which differ only in the part that deals with abstractions and applications (they share transitions rules for algebraic data types and call-by-need updates). The rules of the machine shown in Figure~\ref{fig:s} correspond to the application-related rules of the \ea STG machine: a rule called \rulename{STG\text{-}XXX} here is called \rulename{XXX} in the formulation by Marlow and Peyton Jones~\cite{DBLP:journals/jfp/MarlowJ06}.
The difference is that partial applications in the \Ea STG machine are allocated in the heap. Therefore, there is no need for a separate $\papt$ configuration in STG, but there is a $\papt$ type of heap objects.

\section{Discussion}

Eval/apply machines are also known as `eval/continue' machines (see Danvy~\cite{danvyevalcontinue} for a discussion), since they have a strong connection with continuations and evaluation contexts~\cite{DBLP:journals/tcs/BiernackaD07,citeulike:4128, BRICS-RS-04-26,DBLP:conf/ifl/SieczkowskiBB10}, which makes them modular and more amenable for reasoning. That is why they are attractive underlying evaluation models for modular and formally verified compilation techniques. The direction of our transformation (from \pe to \ea) also reflects the pragmatic choice of \ea over \pe in the Glasgow Haskell Compiler~\cite{DBLP:journals/jfp/MarlowJ06}. Our derivation can be easily applied to different known \pe machines, such as Leroy's ZINC~\cite{Leroy-ZINC}.

Our method is inspired by Danvy \textit{et al.}'s functional correspondence between evaluators and abstract machines~\cite{DBLP:conf/ppdp/AgerBDM03}. It is important to notice that Danvy \textit{et al.}\ use CPS translation as a tool to flatten the structure of evaluators, while here, since the \verb+pop'+ operation is already in a flat, tail-recursive form, we use it to reify the recursive calls of \verb+pop'+ as transitions of the machine, and no new stack of continuations emerges.

The \pe STG machine is in reality a hybrid machine, which is apparent in its original, three-stack formulation~\cite{DBLP:journals/jfp/Jones92a}: the machine is \pe in the argument stack, but it is \ea in the return and update stacks. The equivalence of two simpler variants of the \ea and \pe STG machines has been shown by Encina and Pe{\~n}a~\cite{DBLP:journals/jfp/EncinaP09} by deriving both machines from a single natural semantics. Pir{\'{o}}g and Biernacki~\cite{DBLP:conf/haskell/PirogB10} used a technique based on Danvy's functional correspondence to derive the \Pe STG machine from a big-step operational semantics.

\section*{Acknowledgements}

We are grateful to Filip Sieczkowski, Dariusz Biernacki, and the anonymous reviewers for their valuable comments.


\begin{thebibliography}{10}
\providecommand{\bibitemdeclare}[2]{}
\providecommand{\surnamestart}{}
\providecommand{\surnameend}{}
\providecommand{\urlprefix}{Available at }
\providecommand{\url}[1]{\texttt{#1}}
\providecommand{\href}[2]{\texttt{#2}}
\providecommand{\urlalt}[2]{\href{#1}{#2}}
\providecommand{\doi}[1]{doi:\urlalt{http://dx.doi.org/#1}{#1}}
\providecommand{\bibinfo}[2]{#2}

\bibitemdeclare{inproceedings}{DBLP:conf/ppdp/AgerBDM03}
\bibitem{DBLP:conf/ppdp/AgerBDM03}
\bibinfo{author}{Mads~Sig \surnamestart Ager\surnameend},
  \bibinfo{author}{Dariusz \surnamestart Biernacki\surnameend},
  \bibinfo{author}{Olivier \surnamestart Danvy\surnameend} \&
  \bibinfo{author}{Jan \surnamestart Midtgaard\surnameend}
  (\bibinfo{year}{2003}): \emph{\bibinfo{title}{A functional correspondence
  between evaluators and abstract machines}}.
\newblock In: {\sl \bibinfo{booktitle}{Proceedings of the 5th International
  {ACM} {SIGPLAN} Conference on Principles and Practice of Declarative
  Programming, 27-29 August 2003, Uppsala, Sweden}},
  \bibinfo{publisher}{{ACM}}, pp. \bibinfo{pages}{8--19},
  \doi{10.1145/888251.888254}.

\bibitemdeclare{article}{DBLP:journals/tocl/BiernackaD07}
\bibitem{DBLP:journals/tocl/BiernackaD07}
\bibinfo{author}{Malgorzata \surnamestart Biernacka\surnameend} \&
  \bibinfo{author}{Olivier \surnamestart Danvy\surnameend}
  (\bibinfo{year}{2007}): \emph{\bibinfo{title}{A concrete framework for
  environment machines}}.
\newblock {\sl \bibinfo{journal}{{ACM} Transactions on Computational Logic}}
  \bibinfo{volume}{9}(\bibinfo{number}{1}), pp. \bibinfo{pages}{1--30},
  \doi{10.1145/1297658.1297664}.

\bibitemdeclare{article}{DBLP:journals/tcs/BiernackaD07}
\bibitem{DBLP:journals/tcs/BiernackaD07}
\bibinfo{author}{Malgorzata \surnamestart Biernacka\surnameend} \&
  \bibinfo{author}{Olivier \surnamestart Danvy\surnameend}
  (\bibinfo{year}{2007}): \emph{\bibinfo{title}{A syntactic correspondence
  between context-sensitive calculi and abstract machines}}.
\newblock {\sl \bibinfo{journal}{Theoretical Computer Science}}
  \bibinfo{volume}{375}(\bibinfo{number}{1-3}), pp. \bibinfo{pages}{76--108},
  \doi{10.1016/j.tcs.2006.12.028}.

\bibitemdeclare{inproceedings}{citeulike:4128}
\bibitem{citeulike:4128}
\bibinfo{author}{Olivier \surnamestart Danvy\surnameend}
  (\bibinfo{year}{2004}): \emph{\bibinfo{title}{On Evaluation Contexts,
  Continuations, and the Rest of the Computation}}.
\newblock In \bibinfo{editor}{Hayo \surnamestart Thielecke\surnameend}, editor:
  {\sl \bibinfo{booktitle}{ACM-SIGPLAN Continuations Workshop (CW'04)}},
  \bibinfo{volume}{Technical report CSR-04-1}, \bibinfo{address}{School of
  Computer Science, University of Birmingham, United Kingdom}, pp.
  \bibinfo{pages}{13--23}.
\newblock \urlprefix\url{http://cs.au.dk/~danvy/DSc/29_danvy_cw-2004.pdf}.

\bibitemdeclare{incollection}{danvyevalcontinue}
\bibitem{danvyevalcontinue}
\bibinfo{author}{Olivier \surnamestart Danvy\surnameend}
  (\bibinfo{year}{2009}): \emph{\bibinfo{title}{Towards Compatible and
  Interderivable Semantic Specifications for the {S}cheme Programming Language,
  Part I: Denotational Semantics, Natural Semantics, and Abstract Machines}}.
\newblock In \bibinfo{editor}{Jens \surnamestart Palsberg\surnameend}, editor:
  {\sl \bibinfo{booktitle}{Semantics and Algebraic Specification}}, {\sl
  \bibinfo{series}{Lecture Notes in Computer Science}} \bibinfo{volume}{5700},
  \bibinfo{publisher}{Springer Berlin Heidelberg}, pp.
  \bibinfo{pages}{162--185}, \doi{10.1007/978-3-642-04164-8\_9}.

\bibitemdeclare{article}{DBLP:journals/ipl/DanvyM08}
\bibitem{DBLP:journals/ipl/DanvyM08}
\bibinfo{author}{Olivier \surnamestart Danvy\surnameend} \&
  \bibinfo{author}{Kevin \surnamestart Millikin\surnameend}
  (\bibinfo{year}{2008}): \emph{\bibinfo{title}{On the equivalence between
  small-step and big-step abstract machines: a simple application of
  lightweight fusion}}.
\newblock {\sl \bibinfo{journal}{Information Processing Letters}}
  \bibinfo{volume}{106}(\bibinfo{number}{3}), pp. \bibinfo{pages}{100--109},
  \doi{10.1016/j.ipl.2007.10.010}.

\bibitemdeclare{techreport}{BRICS-RS-04-26}
\bibitem{BRICS-RS-04-26}
\bibinfo{author}{Olivier \surnamestart Danvy\surnameend} \&
  \bibinfo{author}{Lasse~R. \surnamestart Nielsen\surnameend}
  (\bibinfo{year}{2004}): \emph{\bibinfo{title}{Refocusing in Reduction
  Semantics}}.
\newblock \bibinfo{type}{Technical Report} \bibinfo{number}{RS-04-26},
  \bibinfo{institution}{Basic Research in Computer Science (BRICS)},
  \bibinfo{address}{University of Aarhus, Denmark}.
\newblock \urlprefix\url{http://www.brics.dk/RS/04/26/BRICS-RS-04-26.pdf}.

\bibitemdeclare{article}{DBLP:journals/jfp/EncinaP09}
\bibitem{DBLP:journals/jfp/EncinaP09}
\bibinfo{author}{Alberto \surnamestart de~la Encina\surnameend} \&
  \bibinfo{author}{Ricardo \surnamestart Pe{\~{n}}a{-}Mar{\'{\i}}\surnameend}
  (\bibinfo{year}{2009}): \emph{\bibinfo{title}{From natural semantics to {C:}
  {A} formal derivation of two {STG} machines}}.
\newblock {\sl \bibinfo{journal}{Journal of Functional Programming}}
  \bibinfo{volume}{19}(\bibinfo{number}{1}), pp. \bibinfo{pages}{47--94},
  \doi{10.1017/S0956796808006746}.

\bibitemdeclare{article}{DBLP:journals/jfp/Jones92a}
\bibitem{DBLP:journals/jfp/Jones92a}
\bibinfo{author}{Simon L.~Peyton \surnamestart Jones\surnameend}
  (\bibinfo{year}{1992}): \emph{\bibinfo{title}{Implementing Lazy Functional
  Languages on Stock Hardware: The Spineless Tagless G-Machine}}.
\newblock {\sl \bibinfo{journal}{Journal of Functional Programming}}
  \bibinfo{volume}{2}(\bibinfo{number}{2}), pp. \bibinfo{pages}{127--202},
  \doi{10.1017/S0956796800000319}.

\bibitemdeclare{article}{DBLP:journals/lisp/Krivine07}
\bibitem{DBLP:journals/lisp/Krivine07}
\bibinfo{author}{Jean{-}Louis \surnamestart Krivine\surnameend}
  (\bibinfo{year}{2007}): \emph{\bibinfo{title}{A call-by-name lambda-calculus
  machine}}.
\newblock {\sl \bibinfo{journal}{Higher-Order and Symbolic Computation}}
  \bibinfo{volume}{20}(\bibinfo{number}{3}), pp. \bibinfo{pages}{199--207},
  \doi{10.1007/s10990-007-9018-9}.

\bibitemdeclare{techreport}{Leroy-ZINC}
\bibitem{Leroy-ZINC}
\bibinfo{author}{Xavier \surnamestart Leroy\surnameend} (\bibinfo{year}{1990}):
  \emph{\bibinfo{title}{The {ZINC} experiment: an economical implementation of
  the {ML} language}}.
\newblock \bibinfo{type}{Technical report} \bibinfo{number}{117},
  \bibinfo{institution}{INRIA}.
\newblock \urlprefix\url{http://gallium.inria.fr/~xleroy/publi/ZINC.pdf}.

\bibitemdeclare{article}{DBLP:journals/jfp/MarlowJ06}
\bibitem{DBLP:journals/jfp/MarlowJ06}
\bibinfo{author}{Simon \surnamestart Marlow\surnameend} \&
  \bibinfo{author}{Simon L.~Peyton \surnamestart Jones\surnameend}
  (\bibinfo{year}{2006}): \emph{\bibinfo{title}{Making a fast curry: push/enter
  vs. eval/apply for higher-order languages}}.
\newblock {\sl \bibinfo{journal}{Journal of Functional Programming}}
  \bibinfo{volume}{16}(\bibinfo{number}{4-5}), pp. \bibinfo{pages}{415--449},
  \doi{10.1017/S0956796806005995}.

\bibitemdeclare{inproceedings}{DBLP:conf/haskell/PirogB10}
\bibitem{DBLP:conf/haskell/PirogB10}
\bibinfo{author}{Maciej \surnamestart Pir{\'{o}}g\surnameend} \&
  \bibinfo{author}{Dariusz \surnamestart Biernacki\surnameend}
  (\bibinfo{year}{2010}): \emph{\bibinfo{title}{A systematic derivation of the
  {STG} machine verified in Coq}}.
\newblock In \bibinfo{editor}{Jeremy \surnamestart Gibbons\surnameend}, editor:
  {\sl \bibinfo{booktitle}{Proceedings of the 3rd {ACM} {SIGPLAN} Symposium on
  Haskell, Haskell 2010, Baltimore, MD, USA, 30 September 2010}},
  \bibinfo{publisher}{{ACM}}, pp. \bibinfo{pages}{25--36},
  \doi{10.1145/1863523.1863528}.

\bibitemdeclare{inproceedings}{DBLP:conf/ifl/SieczkowskiBB10}
\bibitem{DBLP:conf/ifl/SieczkowskiBB10}
\bibinfo{author}{Filip \surnamestart Sieczkowski\surnameend},
  \bibinfo{author}{Malgorzata \surnamestart Biernacka\surnameend} \&
  \bibinfo{author}{Dariusz \surnamestart Biernacki\surnameend}
  (\bibinfo{year}{2010}): \emph{\bibinfo{title}{Automating Derivations of
  Abstract Machines from Reduction Semantics: {A} Generic Formalization of
  Refocusing in Coq}}.
\newblock In \bibinfo{editor}{Jurriaan \surnamestart Hage\surnameend} \&
  \bibinfo{editor}{Marco~T. \surnamestart Moraz{\'{a}}n\surnameend}, editors:
  {\sl \bibinfo{booktitle}{Implementation and Application of Functional
  Languages---22nd International Symposium, {IFL} 2010, Alphen aan den Rijn,
  The Netherlands, September 1-3, 2010, Revised Selected Papers}}, {\sl
  \bibinfo{series}{Lecture Notes in Computer Science}} \bibinfo{volume}{6647},
  \bibinfo{publisher}{Springer}, pp. \bibinfo{pages}{72--88},
  \doi{10.1007/978-3-642-24276-2\_5}.

\end{thebibliography}
\end{document}